\documentclass{LT23auth}
\usepackage{graphicx}

\begin{document}

\begin{frontmatter}

% Use lower case letters in the title.
\title{Tunneling images of a 2D electron system in a quantizing magnetic field}

\author[address1]{I. J. Maasilta \thanksref{thank1}},
\author[address1]{Subhasish Chakraborty},
\author[address1]{I. Kuljanishvili},
\author[address1]{S. H. Tessmer},
\author[address2]{M. R. Melloch}

\address[address1]{Department of Physics and Astronomy, Michigan State University, East Lansing, Michigan 48824, USA}

\address[address2]{Department of Electrical Engineering, Purdue University, West Lafayette, Indiana 47907, USA}

% The corresponding author should be distinguished and his email
% address and/or fax number must be given. His mailing address has to
% be complete: the proofs are send to this address around
% January 1, 2003. The address for sending proofs has to be indicated
% as "present address", if it is different from the address above.
\thanks[thank1]{Corresponding author. Present address: Department of
Physics, University of Jyv\"askyl\"a, P. O. Box 35, FIN-40351 Jyv\"askyl\"a, Finland 
 E-mail: maasilta@phys.jyu.fi }

\begin{abstract}
We have applied a scanning probe method, Subsurface Charge Accumulation (SCA) imaging, to resolve the local structure of the interior of
 a semiconductor two-dimensional electron system (2DES) in a tunneling geometry. Near magnetic fields corresponding to integer Landau
  level filling,
  submicron scale spatial structure in the out-of-phase component of the tunneling signal becomes visible.
   In the images presented here, the structure repeats itself when the filling factor is changed from $\nu=6$ to $\nu=7$. Therefore, 
   we believe the images  reflect  small modulations in the 2DES density caused by the disorder in the sample. 

\end{abstract}

%
% write here 3 or 4 keywords separated by semicolons
%
\end{frontmatter}

%\section{}

Scanned probe techniques sensitive to electric fields can provide direct images of 2D electron systems \cite{scan}. 
 Here, we present an extension of Subsurface Charge Accumulation (SCA) imaging \cite{sca}, 
 adapted to probe the 2DES interior in a novel tunneling geometry \cite{preprint}. 
 The measurement works as follows and is schematically shown in Fig. 1(a): A tunneling barrier separates the 2D layer and a parallel 
 3D substrate. 
 An ac excitation voltage applied between the substrate and a sharp metal tip locally induces charge to tunnel back and forth 
 between the 3D and 2D layers.  The measured signal is the resulting ac image charge on the tip electrode, and  no charge tunnels 
 directly onto the tip.  In this way, the experiment 
 provides a local measurement of the ability of the 2D system to accommodate additional electrons. 
 
 The sample used for these measurements was an Al$_{0.3}$Ga$_{.7}$As / GaAs (001) wafer grown by molecular beam epitaxy (MBE),
  shown schematically in Fig. 1(a). The 2DES is located a distance of 60 nm below the exposed surface, and a distance 40 nm above a 
  degenerately doped ($10^{18}$ cm$^{-3}$) GaAs substrate (3D metal). The average electron density in the 2DES is $6\times 10^{11}$ 
  cm$^{-2}$,
   the low-temperature transport mobility is $\sim 10^5$ cm$^2$/Vs and the zero magnetic field 3D-2D tunneling rate is 
   approximately 200 kHz.  We position the tip to within a few nanometers of the sample surface using an innovative 
   scanning head that provides a high degree of mechanical and thermal stability \cite{scope}. The image charge signal is detected using 
   a circuit constructed from low-input-capacitance high-electron-mobility transistors. 
    To acquire images, the tip is scanned laterally across the surface without the use of feedback. 
	 The data were acquired with the system immersed in liquid helium-3 at a temperature of 270 mK, with a 20 kHz 
	 applied excitation voltage of 8 mV rms.
	
	%
%  Here is the template to include a figure.
%
\begin{figure}[tbp]
%h=here, t=top, b=bottom, p=separate figure page
\begin{center}\leavevmode
\includegraphics[width=0.9\linewidth]{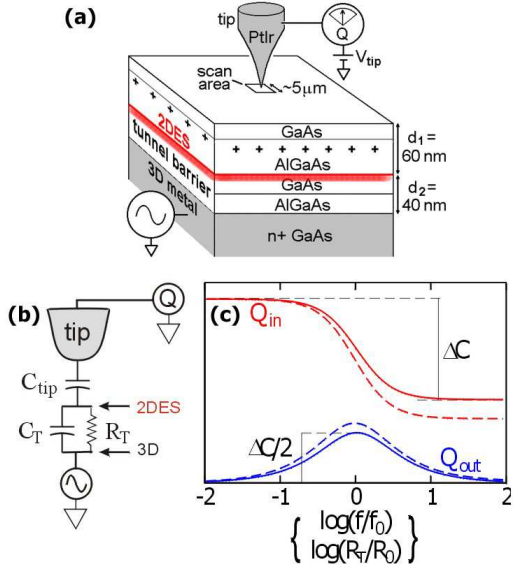}
\caption{(a) Schematic of heterostructure sample and SCA measurement.  The 2DES forms in the potential well 
at the GaAs/AlGaAs interface from electrons provided by silicon dopants (+'s).  
An AlGaAs tunneling barrier separates the 2DES from a 3D substrate. (b) Equivalent circuit of the sample-tip system. 
 (c) In-phase and  90$^\circ$ out-of-phase signals ($Q_{in}$ and $Q_{out}$) versus frequency $f$ and tunneling resistance $R_T$. 
  The charging has identical functional dependence on $f$ and $R_T$ , with characteristic values of  $f_0=[2\pi R_T(C_T+C_{tip})]^{-1}$
   and $R_0=[2\pi f(C_T+C_{tip})]^{-1}$, respectively. 
  The magnitude of the curves is set by the tip-sample capacitance difference $\Delta C$ between a fully charging and locally
   non-charging 2DES. 
  The dashed curves qualitatively show the enhancements in  $Q_{in}$ and $Q_{out}$ as the in-plane relaxation 
  rates approach the tunneling rate.   }
\label{fig1}\end{center}\end{figure}

	Fig. 1(b) shows a simple equivalent circuit for the measurement: The local contribution to the signal is modeled as a tip-2DES 
	capacitance $C_{tip}$ in series with the tunneling barrier, represented as a capacitor $C_T$ and resistor $R_T$ in parallel. 
	To account for phase shifts we define two amplitudes $Q_{in}$ and $Q_{out}$, representing respectively the in-phase and 90$^\circ$
	 out-of-phase (lagging) components of the image charge per unit excitation voltage, typical units being aF. 
	 Fig. 1(c) shows the charging characteristics of the circuit.  
	 In zero magnetic field, the tunneling rate across the barrier $f_0$ is about ten times higher than the 20 kHz excitation frequency. 
	 In this frequency range, the out-of-phase component is more sensitive to variations in the local tunneling resistance.
	 In contrast, the in-phase component is mostly sensitive to capacitance variations. 
	 More rigorous models show an enhancement in the charging characteristics as the 2DES becomes 
	 an increasingly poor conductor, i.e., as the system approaches integer filling, as indicated qualitatively by the dashed 
	 curves in Fig. 1(c).

As the magnetic field approaches integer fillings $\nu=6$ and $\nu=7$, we observe an increase in $Q_{out}$, corresponding to an 
increase in $R_T$, as shown in the fixed-tip signal of Fig. 2(a).  
 Clear and reproducible features appear also in image scans, as exemplified in Fig. 2(b). As one lowers the field the structure appears
 at around $\nu=6$, then disappears and again reappears at around $\nu=7$. This re-emergent behavior likely originates  
 from density modulations induced by the static disorder potential, consistent with the interpretation in \cite{sca}. 
 Structure is seen down to a length scale of few hundred nm; the spatial resolution of our technique is set by the radius of curvature of the tip, 
 nominally 50 nm in this case. More surprisingly, structure also exists in length scales $\sim 1 \mu$m, in contrast to the expextation 
 that the disorder length scale is set by the distance to the donors and to the 3D layer (both 15 nm in our case).

\begin{figure}[tbp]
%h=here, t=top, b=bottom, p=separate figure page
\begin{center}\leavevmode
\includegraphics[width=0.95\linewidth]{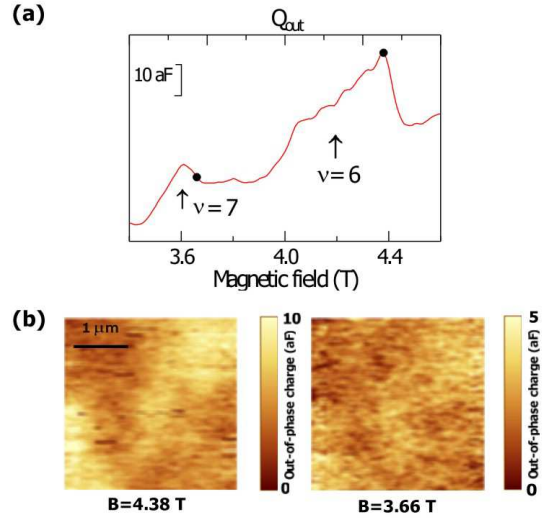}
\caption{Out-of-phase data near $\nu=6$ and $\nu=7$.
 (a) A fixed-tip $Q_{out}$ curve displaying the $\nu=6$ and $\nu=7$ peaks, acquired near the upper right corner of the imaged area. 
 (b) Two $3 \mu$m $\times 3 \mu$m images with values of $B$ indicated as dots in (a).  All data was acquired with the tip bias
  adjusted to equal the contact potential \cite{Yoo}, so that the tip itself
	  does not produce a dc perturbation of the 2DES, as determined by in situ Kelvin probe measurements. Charge sensitivity calculated from this data is
	   $\sim$ 0.06 e/Hz$^{1/2}$. Topographic component of $Q_{out}$ has been subtracted \cite{preprint}.}
\label{fig2}\end{center}\end{figure}

%\section{}
  
%
% for acknowledgement
%
%\begin{ack}
This work was supported by the National Science Foundation (DMR-0075230).
 SHT acknowledges support of the Alfred P. Sloan Foundation.
%\end{ack}

%
% The format of reference should be
% Author1, Author2, Author3, Journal {\bf volume} (year) page.
% No ``and'' between the authors are necessary. 
%

\end{document}